\documentclass[aps,twocolumn,floatfix,pre,showpacs]{revtex4-1}
\usepackage{amsmath}    
\usepackage{graphicx}   
\usepackage{verbatim}   
\usepackage{color}   
\usepackage{subfigure}  
\usepackage{hyperref}   
\usepackage{pstricks}
\usepackage{hyperref}

\begin{document}

\title{Stripe, gossamer, and glassy phases in systems 
with strong non-pairwise interactions}
\author{Karl A. H.  Sellin$^{1}$ and Egor Babaev$^{1,2}$}
\affiliation{$^1$Department of Theoretical Physics, The Royal Institute of
  Technology, SE-10691 Stockholm, Sweden}  
\affiliation{$^2$Department of Physics, University of Massachusetts, Amherst,
  Massachusetts 01003, USA}
\date{\today}
\begin{abstract}
We study structure formation in systems of classical particles in two dimensions with long-range attractive short-range repulsive two-body interactions and repulsive three-body interactions. Stripe, gossamer, and glass phases are found as a result of nonpairwise interaction.
\end{abstract}


\maketitle

\section{Introduction}
The problem of stripe and cluster formation is important in a wide range of physical 
systems, ranging from soft matter \cite{Malescio2003,Glaser2007,
OlsonReichhardt2011}, to hard condensed matter \cite{Spivak2005,
Parameswaran2012} and magnets \cite{Nielsen2008}, to name a few. A case of 
stripe formation which has been especially widely investigated is systems of 
particles with competing multi-scale two-body interactions. In the context of 
superconductivity, a multi-scale long-range attractive short-range repulsive 
interaction is possible between vortices in multicomponent type-1.5 
superconductors \cite{Babaev2005}, for a review see \cite{Babaev20122}. 
The recent experimental claims of stripe and gossamer phases of vortex matter in 
superconductors \cite{Moshchalkov2009,Nishio2010,Gutierrez2012} prompted 
theoretical investigations of whether such structure formation of vortex matter in 
multi-band superconductors is possible or not (see e.g. \cite{Carlstrom2011a,
Dao2011,Drocco2012,geurts}).

Intervortex potentials with short-range repulsive long-range attractive pairwise 
interaction only allow formation of simple clusters in equilibrium situations. In 
\cite{Carlstrom2011a} the question was raised if, in principle, stripe phases can 
occur as a result of non-pairwise intervortex forces. The calculated three-body 
intervortex forces in type-1.5 superconductors are repulsive \cite{Carlstrom2011a,
Edstrom2013} and can certainly be sufficiently strong to result in stripe formation 
for kinetic and entropic reasons. However, since accurate calculations of 
intervortex many-body forces in field theory is highly computationally demanding, 
they have been investigated in only a small number of cases.

Here we ask the following more general question: what kind of unconventional 
ordering patterns can occur in systems with repulsive non-pairwise interactions? 
Previous studies of the structural effects of non-pairwise interactions have shown 
that for a short-range attractive, long-range repulsive pairwise interaction, an 
attractive or repulsive non-pairwise interaction had little effect for the ranges 
studied \citep{Meilhac2011}. In \citep{Kim1999} it was reported that repulsive
pairwise and attractive non-pairwise interactions have been found to cause 
clustering of particles under certain conditions. The works 
\citep{Sengupta2007,Sengupta2010} simulated driven crystal phases in 
two-dimensional systems with three-body forces. We investigate a model 
of point particles, with long-range attractive short-range repulsive two-body 
interaction, and repulsive three-body interaction. We will investigate the structure 
formation of such a system by tuning the relative strength of the two- and three-body 
interactions, as well as temperature and particle density. We will demonstrate that 
the system possesses a rich variety of pattern formation such  as stripe, gossamer 
and glassy phases.

%%%%%%%%%%%%%%%%%%% MODEL DESCRIPTION
\section{Model}

Consider particles interacting with a pairwise potential which is repulsive at short 
particle separation and attractive at longer separation, such that there is a 
preferred separation between two particles. In the case of three particles with such 
a pairwise interaction, the ground state configuration will occur when the particles 
form an equilateral triangle, with a line constituting an energetically excited state. 
In the case of many particles, the tendency of three particles to form a triangle will 
favor a hexagonal symmetry of the structure formation. In this paper we will begin 
by considering how the ground state of three particles is changed by adding a 
repulsive non-pairwise interaction upon the two-body interaction and we will then 
show how the non-pairwise interaction affects the structure formation in systems 
of many particles. As we will see, the ground state of three particles will for a 
sufficiently strong three-body repulsion, be that of a straight line instead of a 
triangular configuration, a tendency which will cause a variety of structural phases 
in systems of many particles.

\begin{figure}
	\includegraphics[width=0.95\columnwidth]{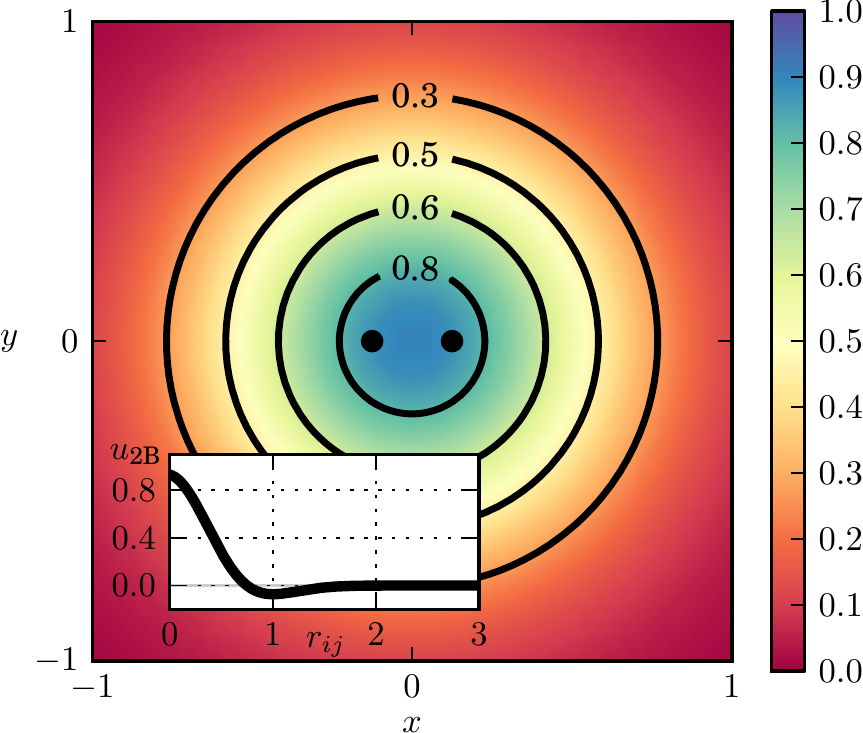}
	\caption{(Color online) Depiction of the model studied. The contour plot is of 
	the three-body interaction energy experienced by a particle with position 
	$\boldsymbol{r}=(x,y)$ by two other particles fixed at positions 
	$\boldsymbol{r}_1=(-0.125,0)$, $\boldsymbol{r}_2=(0.125,0)$ 
	(indicated by black dots), i.e. a plot of $f(\boldsymbol{r},\boldsymbol{r}_1,
	\boldsymbol{r}_2)$ with $f$ defined in Eq. \eqref{eq:f}. The model parameters
	 are $\varepsilon_{\mbox{\scriptsize 3B}}=1.0$, $\alpha=2.0$, $w=0.8$. 
	The inset shows the two-body interaction potential 
	\eqref{eq:two_body_interaction_potential} against particle separation with 
	$\varepsilon_{\mbox{\scriptsize 2B}}=1.0$, $a=3.0$, $b=0.2$, $c=3.0$ 
	and $d=0.60$.}
	\label{fig:potentials_plot}
\end{figure}
The total interaction potential energy $U(\boldsymbol{X})$ of $N$ classical point 
particles with two-body and three-body interactions in the state 
$\boldsymbol{X}=\{\boldsymbol{r}_1,\boldsymbol{r}_2,
\dots,\boldsymbol{r}_N\}$, is in general given by
\begin{align}
	U(\boldsymbol{X})=&
	\sum_{i=1}^N\sum_{j=i+1}^Nu_{\mbox{\scriptsize 2B}}(\boldsymbol{r}_i,
	\boldsymbol{r}_j)+\nonumber\\
	+&\sum_{i=1}^N\sum_{j=i+1}^N\sum_{k=j+1}^Nu_{\mbox{\scriptsize 3B}}
	(\boldsymbol{r}_{i},\boldsymbol{r}_{j},\boldsymbol{r}_{k}),
	\label{eq:total_potential_energy}
\end{align}	
where $u_{\mbox{\scriptsize 2B}}$ corresponds to the pairwise two-body 
interaction (which we take to be long-range attractive and short-range repulsive) 
and $u_{\mbox{\scriptsize 3B}}$ corresponds to the non-pairwise three-body 
interaction (taken to be purely repulsive). We model the pairwise interaction 
$u_{\mbox{\scriptsize 2B}}$ of two particles $i,j$ as a sum of Gaussians
\begin{equation}
	\frac{u_{\mbox{\scriptsize 2B}}(r_{ij})}{\varepsilon_{\mbox{\scriptsize 2B}}}
	=
	\mathrm{e}^{-a r_{ij}^2}-b\mathrm{e}^{-c\left(r_{ij}-d\right)^2},
	\label{eq:two_body_interaction_potential}
\end{equation}
where $\varepsilon_{\mbox{\scriptsize 2B}}$ is a parameter that determines the
 strength of the interaction, and 
$r_{ij}=\vert\boldsymbol{r}_i-\boldsymbol{r}_j\vert$ is the distance between 
particles $i$ and $j$.

The three-body interaction potential of three particles $i,j,k$ is modeled by
\begin{align}
	\frac{u_{\mbox{\scriptsize 3B}}(\boldsymbol{r}_{i},\boldsymbol{r}_{j},
	\boldsymbol{r}_{k})}{\varepsilon_{\mbox{\scriptsize 3B}}}
	=
	f(\boldsymbol{r}_i,&\boldsymbol{r}_j,\boldsymbol{r}_k)+\nonumber
	\\+&f(\boldsymbol{r}_j,\boldsymbol{r}_i,\boldsymbol{r}_k)+
	f(\boldsymbol{r}_k,\boldsymbol{r}_i,\boldsymbol{r}_j),
	\label{eq:three_body_interaction_potential}
\end{align}
where the function $f$ is a two-dimensional Gaussian
\begin{align}
	f(\boldsymbol{r}_i,\boldsymbol{r}_j,\boldsymbol{r}_k)
	=
	&\mathrm{e}^{-\alpha\left((x_i-R_x)^2+(y_i-R_y)^2\right)-\ell^2/w^2},
	\label{eq:f}
\end{align}
where $\boldsymbol{R}=(R_x,R_y)=(\boldsymbol{r}_j+\boldsymbol{r}_k)/2$ is the 
center-of-mass of the pair ($j,k$), 
$\ell=\vert\boldsymbol{r}_j-\boldsymbol{r}_k\vert$ 
is the distance between the particles $(j,k)$, and $\alpha$, $w$ are model 
parameters which characterizes the range of the interaction. In Fig. 
\ref{fig:potentials_plot} we plot the potentials for a set of parameters that will 
unless otherwise stated be used throughout this article. These potentials have a 
quite similar form as multi-band intervortex potentials in type-1.5 superconductors 
\cite{Carlstrom2011a,Edstrom2013}.

%%%%%%%%%%%%%%%%%%%%% SIMULATION METHOD
\section{Simulation method}
We investigate structure formation of the system by using the Metropolis Monte 
Carlo (MC) algorithm \cite{Metropolis1949} with parallel tempering 
\cite{Swendsen1986,Earl2005}. Our system is considered to be a fixed number $N$ 
particles inside a $L\times L$ box so that the density $\rho=N/L^2$. We impose 
periodic boundary conditions by the minimum image convention 
\cite{allen1991computer}. We take a MC trial move to be a displacement of a 
randomly chosen particle by a randomly chosen distance in a random direction.

In order to quantitatively assess  the tendency of the system to form a stripe phase, 
we define the parameter
\begin{equation}
 	\Psi_{\mbox{\scriptsize S}}=
	\Bigg\vert-1+\frac{1}{N}\sum_{i=1}^N\Big\vert\sum_{j=1}^2
	\exp (\mathrm{i}2\phi_{ij})\Big\vert\Bigg\vert,
	\label{eq:PsiS}
\end{equation}
where the sum in $j$ runs over the two nearest neighbors of particle $i$, and 
$\phi_{ij}$ is the angle of the line joining the particles, with respect to an 
arbitrary axis. The parameter is constructed such that it is unity if three particles 
form a straight line, and vanishes if the particles form an equilateral triangle. 
For many particles, $\Psi_{\mbox{\scriptsize S}}$ is unity if they form  several 
straight lines, or close to unity if they form curved and/or intersecting lines.

%%%%%%%%%%%%%%%%%%%%%%% RESULTS
\section{Results}
\begin{figure}
	\includegraphics[width=0.95\columnwidth]{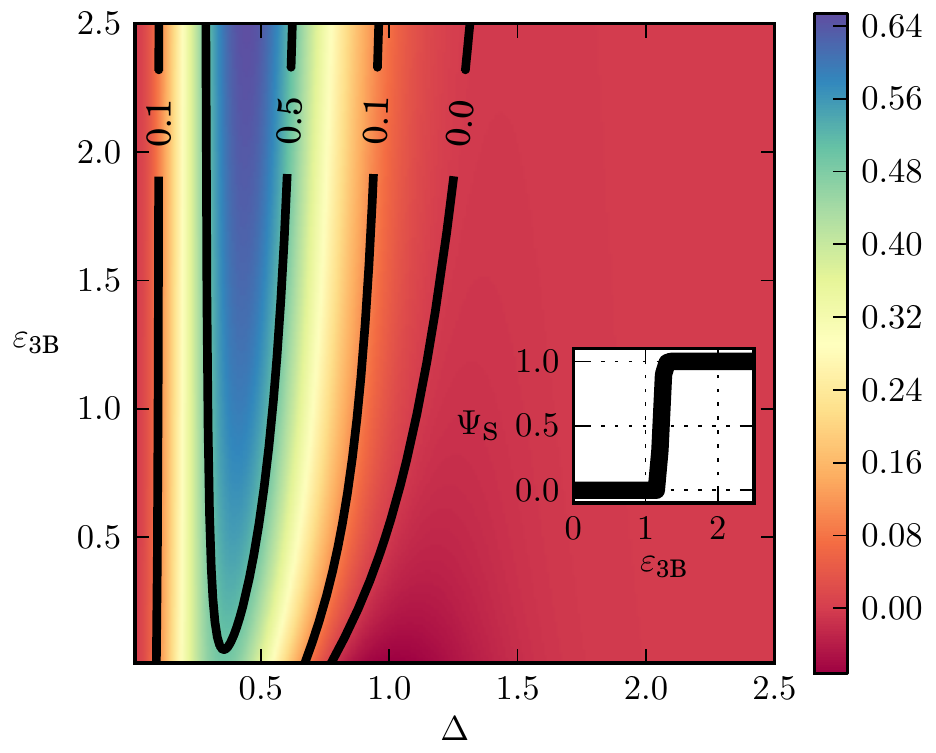}
	\caption{(Color online) The existence of a stripe phase for $N=3$. Displayed 
	is the difference $\Delta E$ of an equilateral triangle of side $\Delta$ and a 
	straight equidistant line with spacing $\Delta$, as a function of $\Delta$ and 
	$\varepsilon_{\mbox{\scriptsize 3B}}$. In the region to the left, where 
	$\Delta E>0$, the three-body repulsion is sufficiently strong to cause the 
	system to energetically favor a line over a triangle. The inset confirms this by 
	showing the ground state value of $\Psi_{\mbox{\scriptsize S}}$ versus 
	$\varepsilon_{\mbox{\scriptsize 3B}}$ obtained from MC simulations with 
	three particles, where the ground state is obtained by slow cooling.}
	\label{fig:triplet_results}
\end{figure} 
We begin by considering the simplest relevant case of a particle triplet, $N=3$, and 
demonstrate that at a certain critical strength of the three-body interaction, it 
becomes energetically favorable for the particles to align in a straight equidistant 
line rather than an equilateral triangle. We compare the two cases by computing the 
total interaction energy of an equilateral triangle with sides $\Delta$, as well as for 
a straight equidistant line with spacing $\Delta$. We plot the difference 
$\Delta E=(E_{\mbox{\scriptsize triangle}}-E_{\mbox{\scriptsize line}})/
(\varepsilon_{\mbox{\scriptsize 2B}}+\varepsilon_{\mbox{\scriptsize 2B}})$ 
as a function of particle spacing 
$\Delta$ and $\varepsilon_{\mbox{\scriptsize 3B}}$, shown in Fig. 
\ref{fig:triplet_results}. As is seen there is a region in which the line configuration 
is energetically favorable,  which is also confirmed with MC simulation shown in the 
inset of Fig. \ref{fig:triplet_results}. As can be checked from Eq. \ref{eq:PsiS}, for 
$N=3$ the parameter $\Psi_{\mbox{\scriptsize S}}$ is exactly unity for a straight 
line, and exactly zero for the case of an equilateral triangle. In the MC simulation 
three particles were given random initial positions and their ground state 
configuration was determined by slowly cooling the system to $T=0$. The ground 
state value of $\Psi_{\mbox{\scriptsize S}}$ is then calculated from the final $T=0$ 
configuration. For each $\varepsilon_{\mbox{\scriptsize 3B}}$, the cooling 
simulation was repeated several times from which the average 
$\Psi_{\mbox{\scriptsize S}}$ was computed.

\begin{figure}
	\includegraphics[width=0.95\columnwidth]{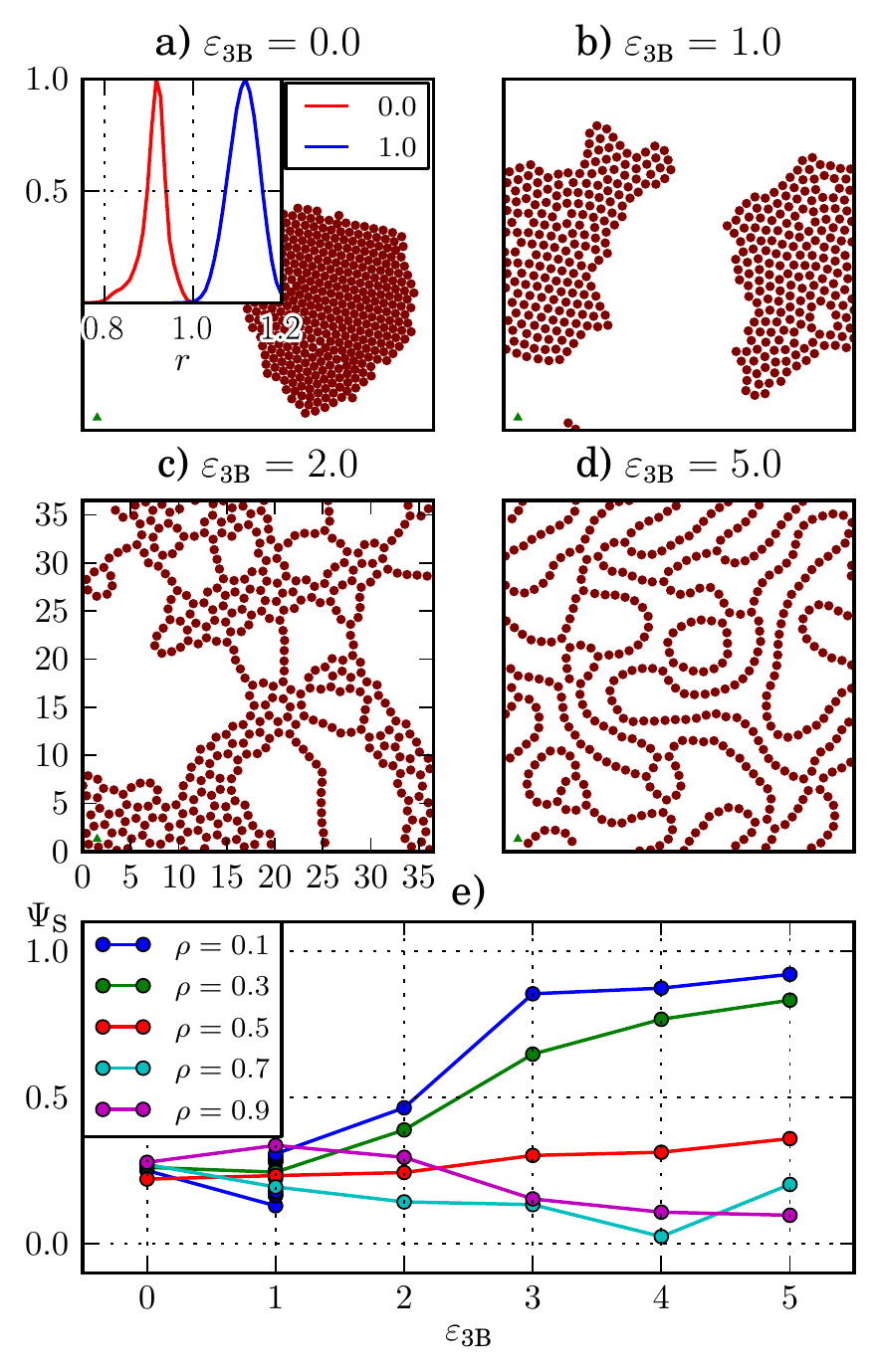}	
	\caption{(Color online) By increasing the three-body interaction strength 
	$\varepsilon_{\mbox{\scriptsize 3B}}$, various phases are induced in systems 
	of many particles. Displayed in a)-d) are structure formation for increasing 
	$\varepsilon_{\mbox{\scriptsize 3B}}$ at constant density $\rho=0.3$ with 
	$N=400$. Panel e) shows how $\Psi_{\mbox{\scriptsize S}}$ depends on the 
	strength of the three-body interaction for various densities. The inset of a) 
	shows normalized distributions of nearest neighbor distances for the cases 
	$\varepsilon_{\mbox{\scriptsize 3B}}=0.0$ and 
	$\varepsilon_{\mbox{\scriptsize 3B}}=1.0$. For comparison, the green 
	triangle in the lower left corner of a)-d) has side equal to the minimum 
	separation of the two-body potential.}
	\label{fig:varying_E3B_low_rho}
\end{figure} 
We now proceed to consider larger systems and the effect of the strength of the 
three-body interaction. MC simulation snapshots for several values of 
$\varepsilon_{\mbox{\scriptsize 3B}}$ are shown in Fig. 
\ref{fig:varying_E3B_low_rho} where the same transition into a stripe phase as in 
the three particle case of Fig. \ref{fig:triplet_results} is seen (see Fig 
\ref{fig:varying_E3B_low_rho} e)). As one increases 
$\varepsilon_{\mbox{\scriptsize 3B}}$ from zero, at first the main effect is to 
increase the mean nearest neighbor distance of the particles (see the inset in Fig. 
\ref{fig:varying_E3B_low_rho} a)), amounting to only a quantitative and not 
qualitative difference in the structure formation, as the preferred number of 
nearest neighbors of a given particle is still six. For the values 
$\varepsilon_{\mbox{\scriptsize 3B}}=0.0,1.0$ (Fig. \ref{fig:varying_E3B_low_rho}
 a) and b)), the two-body interaction dominates, and enforces a hexagonal symmetry. 
However, when $\varepsilon_{\mbox{\scriptsize 3B}}$ surpasses a critical value 
predicted by the results of Fig. \ref{fig:triplet_results}, there is a qualitative change 
as the system will first form a gossamer structure for 
$\varepsilon_{\mbox{\scriptsize 3B}}=2.0$ (Fig. \ref{fig:varying_E3B_low_rho} c)), 
where the system prefers particle bonds with only three nearest neighbors due to a 
competition of the two- and three-body interactions. Further increasing the 
non-pairwise repulsion to $\varepsilon_{\mbox{\scriptsize 3B}}=5.0$ 
(Fig. \ref{fig:varying_E3B_low_rho} d)), a filamentary stripe structure formation 
appears as the system prefers having only two nearest neighbors due to a 
domination of the three-body interaction. We also note in Fig. 
\ref{fig:varying_E3B_low_rho} e) that the stripe phase can only arise at low 
densities where there is room for the particles to spread into their filamentary 
structures. 

\begin{figure}
	\includegraphics[width=0.95\columnwidth]{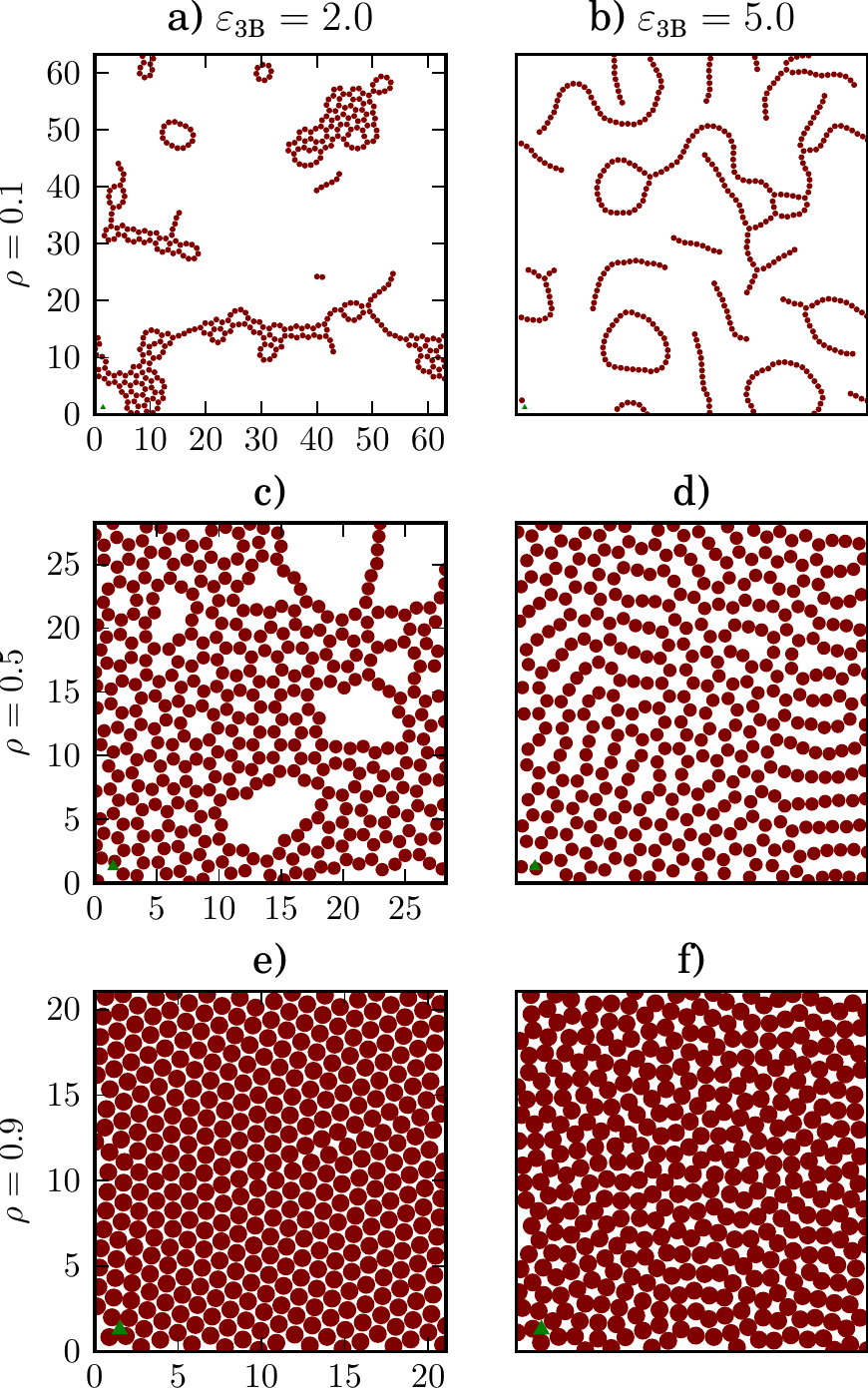}
	\caption{(Color online) By increasing density for the case of competing 
	two-body and three-body interactions (left column) and a dominating 
	three-body interaction (right column), various structural phases are obtained. 
	Here $N=400$.}
	\label{fig:weak_and_strong_E3B_varying_rho}
\end{figure} 
Consider now the effects of increasing density for the cases 
$\varepsilon_{\mbox{\scriptsize 3B}}=2.0$ (where there is a competition between 
the two-body interaction and the three-body interaction, see Fig. 
\ref{fig:varying_E3B_low_rho} c)), and $\varepsilon_{\mbox{\scriptsize 3B}}=5.0$ 
(where the three-body interaction dominates, see Fig. 
\ref{fig:varying_E3B_low_rho} d)). Results are given in Fig. 
\ref{fig:weak_and_strong_E3B_varying_rho}. For the case 
$\varepsilon_{\mbox{\scriptsize 3B}}=2.0$ (left column) the gossamer-like clusters 
will when increasing density be pushed together into a hexagonal lattice. For the 
case $\varepsilon_{\mbox{\scriptsize 3B}}=5.0$ (right column) the filamentary 
stripe structures will first be squeezed into a gossamer structure, where the 
pressure forces some particles to accept having three nearest neighbors rather 
than the preferred value of two. Further increasing the density enhances this 
frustration and creates a disordered state, a tendency we will investigate in the 
next paragraph.

\begin{figure}
	\includegraphics[width=0.95\columnwidth]{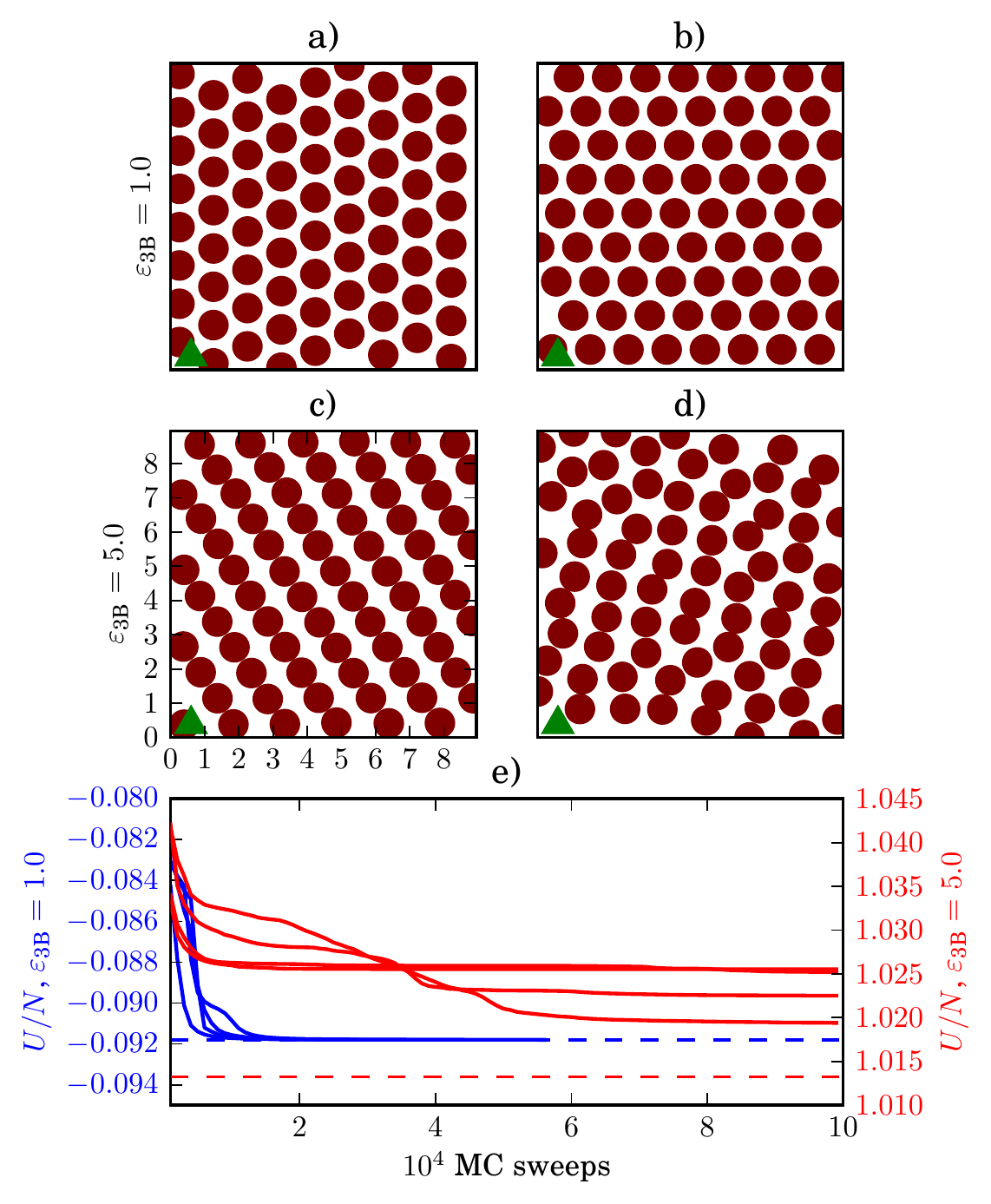}
	\caption{(Color online) A glassy phase occurs for a strong three-body repulsion 
	at high density. Here $\rho=0.9$ with $N=72$. Panels a) and b) show 
	configurations for $\varepsilon_{\mbox{\scriptsize 3B}}=1.0$ obtained by 
	parallel tempering and fast cooling, respectively. Panels c) and d) are for 
	$\varepsilon_{\mbox{\scriptsize 3B}}=5.0$. Panel e) shows the evolution of 
	the total internal energy per particle during four fast cooling simulations for 
	each system, with ground state energies obtained by parallel tempering shown 
	with dashed lines.}
	\label{fig:glass_phase}
\end{figure}
By comparing Fig. \ref{fig:weak_and_strong_E3B_varying_rho} e) and f), it is 
evident that at high densities a relatively weak three-body repulsion yields a 
symmetric lattice and a strong three-body interaction a structurally disordered 
state. This suggests that a strong non-pairwise repulsion creates a glassy phase at 
high densities in the sense that the system is very unlikely to find its ground state 
during a fast cooling \cite{Debenedetti2001}. We investigate this by performing 
long simulations with parallel tempering to find a structurally symmetric ground 
state, and compare with states obtained from $T=0$ MC simulation from a random 
initial configuration (which we consider to be a fast cooling). Results are given in 
Fig. \ref{fig:glass_phase}. As is seen in panel e), the system converges to the 
ground state in the case of fast cooling for 
$\varepsilon_{\mbox{\scriptsize 3B}}=1.0$, indicating a non-glassy system. 
However, for $\varepsilon_{\mbox{\scriptsize 3B}}=5.0$, fast cooling simulations 
consistently fail to produce the ground state. Thus a glassy phase occurs for strong 
non-pairwise repulsion at high densities, as the particles experience frustration 
preventing them to find their ground state, which enforces disorder in the structure 
formation of the system.  

\begin{figure}
	\includegraphics[width=0.95\columnwidth]
	{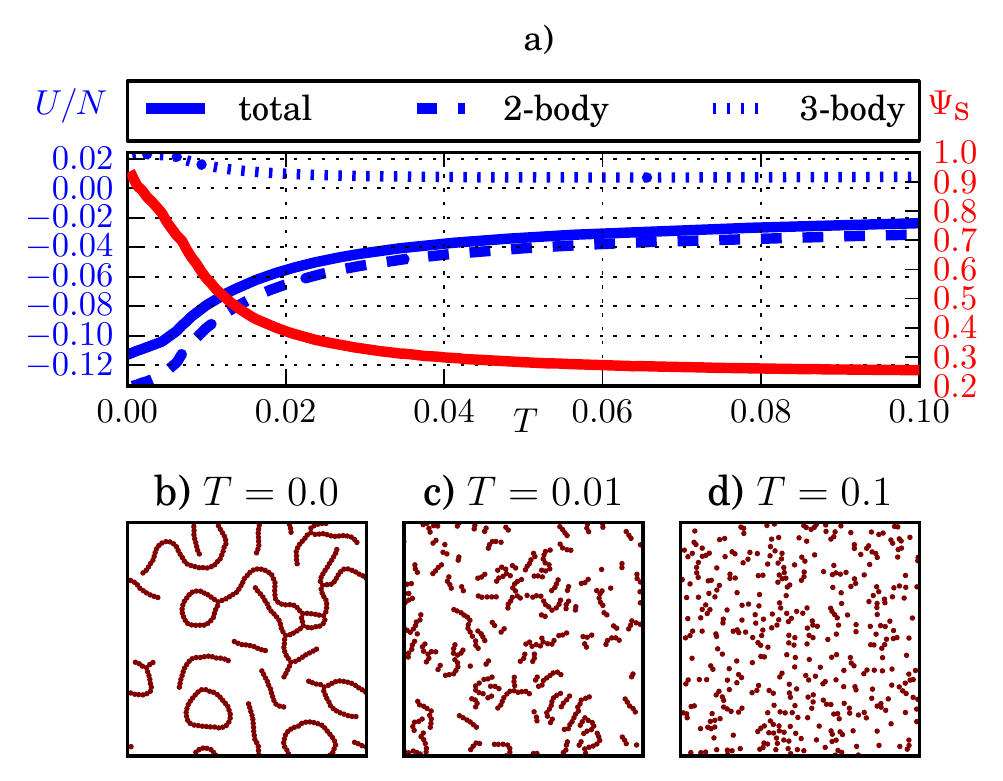}
	\caption{(Color online) Thermal effects of the stripe phase. Here 
	$\varepsilon_{\mbox{\scriptsize 3B}}=5.0$, $\rho=0.1$ with $N=400$. 
	Displayed in a) are the thermally averaged total internal energy $U$ per 
	particle $N$ (with two- and three-body contributions as dashed and dotted 
	lines respectively) and $\Psi_{\mbox{\scriptsize S}}$ against temperature $T$. 
	The lower panels are three snapshots of the system in the low, intermediate, 
	and high temperature phases.}
	\label{fig:stripe_phase_thermal_effects}
\end{figure}	
Next, we consider melting properties of the stripe phase where the system is dilute 
and has strong three-body interactions. As is seen in Fig. 
\ref{fig:stripe_phase_thermal_effects}, the system undergoes a  melting transition 
as temperature is increased, associated with the loss of stripe ordering quantified 
by $\Psi_{\mbox{\scriptsize S}}$. When increasing temperature, individual 
particles can dissociate from their stripes, causing the chains to be broken and 
shorter on average, as seen in Fig. \ref{fig:stripe_phase_thermal_effects} c), 
a process which continues until the system is melted, see Fig. 
\ref{fig:stripe_phase_thermal_effects} d). 
In the melted  phase the three-body contribution to the total interaction energy 
diminishes and almost vanishes, see Fig. \ref{fig:stripe_phase_thermal_effects} a).

\begin{figure}
	\includegraphics[width=0.95\columnwidth]{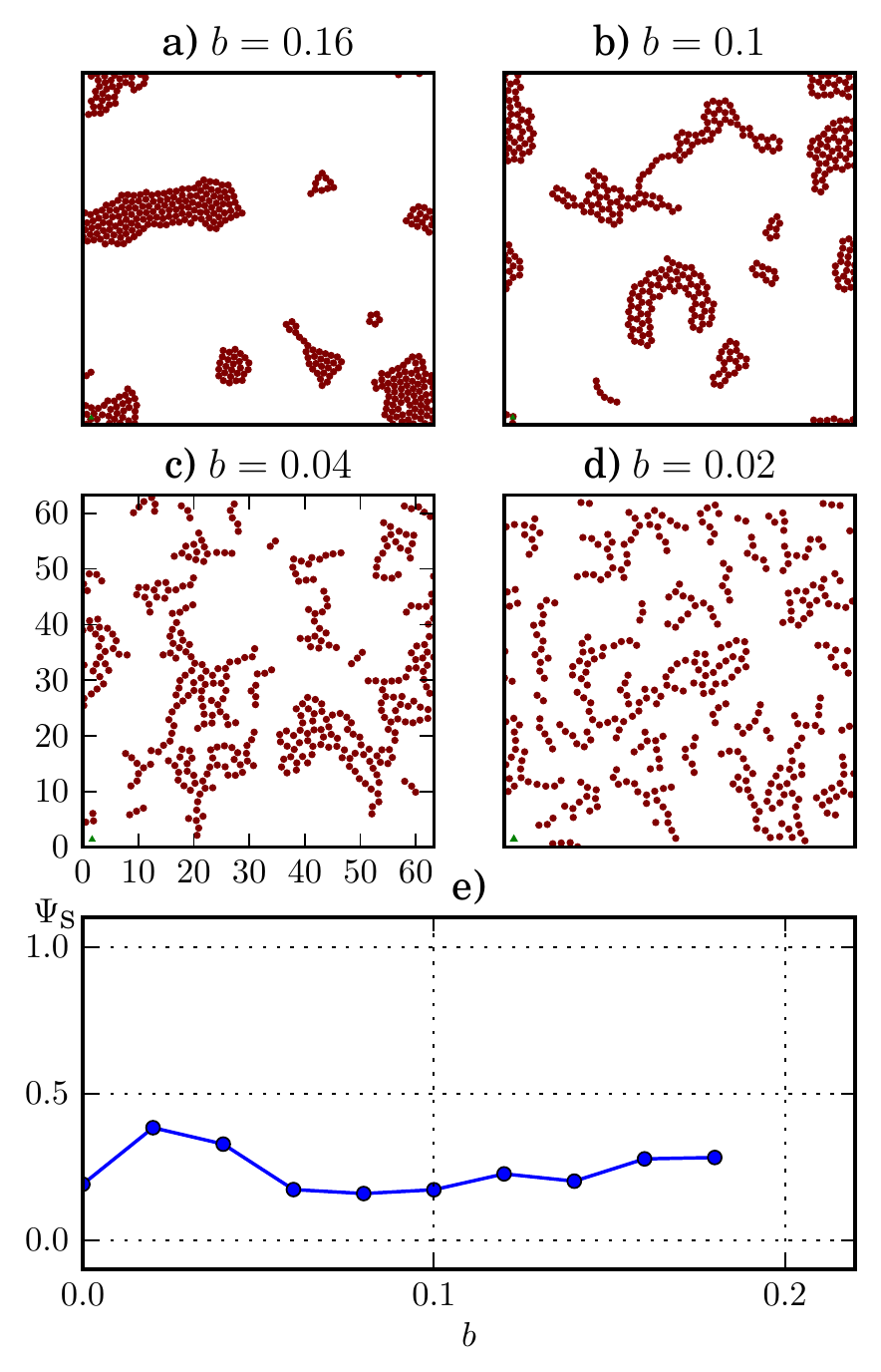}
	\caption{(Color online) Decreasing the two-body minimum by decreasing the 
	parameter $b$ does not induce a clear stripe phase for the parameters studied. Panels a)-d) shows structure 
	formation for $\rho=0.1$, $N=400$, 
	$\varepsilon_{\mbox{\scriptsize 3B}}=1.0$, when decreasing $b$. Panel e) 
	shows how $\Psi_{\mbox{\scriptsize S}}$ depends on $b$.}
	\label{fig:varying_b_low_rho}
\end{figure}
Finally, rather than varying the strength of the three-body interaction 
$\varepsilon_{\mbox{\scriptsize 3B}}$ we consider decreasing the depth of the 
minimum of the two-body potential characterized by the parameter $b$ in Eq. 
\eqref{eq:two_body_interaction_potential}, while keeping the strength of the 
three-body interaction constant. Results are given in Fig.
 \ref{fig:varying_b_low_rho}. Decreasing the minimum $b$ amounts to a weaker 
long-range attraction of the particles and can change the structure formation by 
creating voids as is seen in Fig. \ref{fig:varying_b_low_rho} b), or stripe-like 
tendencies seen in Fig. \ref{fig:varying_b_low_rho} b), c) and d). However, no clear
 stripe phase (compare with Fig. \ref{fig:weak_and_strong_E3B_varying_rho} b)) 
with a significantly high value of $\Psi_{\mbox{\scriptsize S}}$ occurs, as seen in 
Fig. \ref{fig:varying_b_low_rho} e). In the stripe phase, the two-body interaction is 
what binds the particles in the chains, which means that as the long-range attraction 
of the two-body interaction weakens, the particles in the chains become less tightly 
bound, which in turn counteracts the tendency of the three-body interaction to cause 
a stripe phase.
 
%%%%%%%%%%%%%%%%%%%%%%%CONCLUSION & DISCUSSION
\section{Conclusions}
In conclusion, we have demonstrated that for a system of classical particles in two 
dimensions with a two-body short-range repulsive long-range attractive interaction 
favoring clustering with hexagonal symmetry, an additional repulsive three-body 
interaction can significantly alter the structure formation. The form of the potentials 
which we investigated  is inspired by the form of interaction between vortices in 
type-1.5 superconductors. For weak three-body repulsions, the difference is only 
quantitative as the mean nearest neighbor separation of the particles becomes 
larger, but a sufficiently strong three-body repulsion can cause a qualitative change 
as the system enters new phases. In the stripe phase, the ground state of a triplet is 
a straight equidistant line rather than an equilateral triangle, a tendency which can 
be quantified by a parameter $\Psi_{\mbox{\scriptsize S}}$ defined in Eq. 
\eqref{eq:PsiS}. When varying the relative strengths of the pairwise and 
non-pairwise potentials, three phases are found, one where the pairwise interaction 
dominates which yields hexagonal symmetry in the structure formation, one where 
the non-pairwise interaction dominates which yields a stripe phase, and one phase 
where there is competition between the pairwise and non-pairwise interactions, 
which yields a phase of gossamer structure formation. At high densities, a strong 
non-pairwise interaction causes glassy behavior of the system as the particles 
experience frustration and will not easily find their ground state.

%%%%%%%%%%%%%%%%%%%%%%%ACKNOWLEDGMENTS
\section{Acknowledgements}
We thank J. Machta and C. Santangelo for discussions.
The work was supported by Knut and Alice Wallenberg Foundation through the 
Royal Swedish Academy of Sciences Fellowship, Swedish
Research Council and by the National Science Foundation CAREER Award No. 
DMR-0955902.
The computations were performed on resources provided by the Swedish 
National Infrastructure for Computing (SNIC) at National Supercomputer Center 
at Link\"oping, Sweden.

\bibliographystyle{apsrev4-1}
\bibliography{referencesv2}

%merlin.mbs apsrev4-1.bst 2010-07-25 4.21a (PWD, AO, DPC) hacked
%Control: key (0)
%Control: author (72) initials jnrlst
%Control: editor formatted (1) identically to author
%Control: production of article title (-1) disabled
%Control: page (0) single
%Control: year (1) truncated
%Control: production of eprint (0) enabled
\begin{thebibliography}{25}%
\makeatletter
\providecommand \@ifxundefined [1]{%
 \@ifx{#1\undefined}
}%
\providecommand \@ifnum [1]{%
 \ifnum #1\expandafter \@firstoftwo
 \else \expandafter \@secondoftwo
 \fi
}%
\providecommand \@ifx [1]{%
 \ifx #1\expandafter \@firstoftwo
 \else \expandafter \@secondoftwo
 \fi
}%
\providecommand \natexlab [1]{#1}%
\providecommand \enquote  [1]{``#1''}%
\providecommand \bibnamefont  [1]{#1}%
\providecommand \bibfnamefont [1]{#1}%
\providecommand \citenamefont [1]{#1}%
\providecommand \href@noop [0]{\@secondoftwo}%
\providecommand \href [0]{\begingroup \@sanitize@url \@href}%
\providecommand \@href[1]{\@@startlink{#1}\@@href}%
\providecommand \@@href[1]{\endgroup#1\@@endlink}%
\providecommand \@sanitize@url [0]{\catcode `\\12\catcode `\$12\catcode
  `\&12\catcode `\#12\catcode `\^12\catcode `\_12\catcode `\%12\relax}%
\providecommand \@@startlink[1]{}%
\providecommand \@@endlink[0]{}%
\providecommand \url  [0]{\begingroup\@sanitize@url \@url }%
\providecommand \@url [1]{\endgroup\@href {#1}{\urlprefix }}%
\providecommand \urlprefix  [0]{URL }%
\providecommand \Eprint [0]{\href }%
\providecommand \doibase [0]{http://dx.doi.org/}%
\providecommand \selectlanguage [0]{\@gobble}%
\providecommand \bibinfo  [0]{\@secondoftwo}%
\providecommand \bibfield  [0]{\@secondoftwo}%
\providecommand \translation [1]{[#1]}%
\providecommand \BibitemOpen [0]{}%
\providecommand \bibitemStop [0]{}%
\providecommand \bibitemNoStop [0]{.\EOS\space}%
\providecommand \EOS [0]{\spacefactor3000\relax}%
\providecommand \BibitemShut  [1]{\csname bibitem#1\endcsname}%
\let\auto@bib@innerbib\@empty
%</preamble>
\bibitem [{\citenamefont {Malescio}\ and\ \citenamefont
  {Pellicane}(2003)}]{Malescio2003}%
  \BibitemOpen
  \bibfield  {author} {\bibinfo {author} {\bibfnamefont {G.}~\bibnamefont
  {Malescio}}\ and\ \bibinfo {author} {\bibfnamefont {G.}~\bibnamefont
  {Pellicane}},\ }\href {\doibase 10.1038/nmat820} {\bibfield  {journal}
  {\bibinfo  {journal} {Nat. Mater.}\ }\textbf {\bibinfo {volume} {2}},\
  \bibinfo {pages} {97} (\bibinfo {year} {2003})}\BibitemShut {NoStop}%
\bibitem [{\citenamefont {Glaser}\ \emph {et~al.}(2007)\citenamefont {Glaser},
  \citenamefont {Grason}, \citenamefont {Kamien}, \citenamefont {Ko\v{s}mrlj},
  \citenamefont {Santangelo},\ and\ \citenamefont {Ziherl}}]{Glaser2007}%
  \BibitemOpen
  \bibfield  {author} {\bibinfo {author} {\bibfnamefont {M.~A.}\ \bibnamefont
  {Glaser}}, \bibinfo {author} {\bibfnamefont {G.~M.}\ \bibnamefont {Grason}},
  \bibinfo {author} {\bibfnamefont {R.~D.}\ \bibnamefont {Kamien}}, \bibinfo
  {author} {\bibfnamefont {A.}~\bibnamefont {Ko\v{s}mrlj}}, \bibinfo {author}
  {\bibfnamefont {C.~D.}\ \bibnamefont {Santangelo}}, \ and\ \bibinfo {author}
  {\bibfnamefont {P.}~\bibnamefont {Ziherl}},\ }\href {\doibase
  10.1209/0295-5075/78/46004} {\bibfield  {journal} {\bibinfo  {journal}
  {Europhys. Lett.}\ }\textbf {\bibinfo {volume} {78}},\ \bibinfo {pages}
  {46004} (\bibinfo {year} {2007})}\BibitemShut {NoStop}%
\bibitem [{\citenamefont {Olson~Reichhardt}\ \emph {et~al.}(2011)\citenamefont
  {Olson~Reichhardt}, \citenamefont {Reichhardt},\ and\ \citenamefont
  {Bishop}}]{OlsonReichhardt2011}%
  \BibitemOpen
  \bibfield  {author} {\bibinfo {author} {\bibfnamefont {C.~J.}\ \bibnamefont
  {Olson~Reichhardt}}, \bibinfo {author} {\bibfnamefont {C.}~\bibnamefont
  {Reichhardt}}, \ and\ \bibinfo {author} {\bibfnamefont {A.~R.}\ \bibnamefont
  {Bishop}},\ }\href {\doibase 10.1103/PhysRevE.83.041501} {\bibfield
  {journal} {\bibinfo  {journal} {Phys. Rev. E}\ }\textbf {\bibinfo {volume}
  {83}},\ \bibinfo {pages} {041501} (\bibinfo {year} {2011})}\BibitemShut
  {NoStop}%
\bibitem [{\citenamefont {Spivak}\ and\ \citenamefont
  {Kivelson}(2005)}]{Spivak2005}%
  \BibitemOpen
  \bibfield  {author} {\bibinfo {author} {\bibfnamefont {B.}~\bibnamefont
  {Spivak}}\ and\ \bibinfo {author} {\bibfnamefont {S.~A.}\ \bibnamefont
  {Kivelson}},\ }\href {\doibase 10.1103/PhysRevB.72.045355} {\bibfield
  {journal} {\bibinfo  {journal} {Phys. Rev. B}\ }\textbf {\bibinfo {volume}
  {72}},\ \bibinfo {pages} {045355} (\bibinfo {year} {2005})}\BibitemShut
  {NoStop}%
\bibitem [{\citenamefont {Parameswaran}\ \emph {et~al.}(2012)\citenamefont
  {Parameswaran}, \citenamefont {Kivelson}, \citenamefont {Rezayi},
  \citenamefont {Simon}, \citenamefont {Sondhi},\ and\ \citenamefont
  {Spivak}}]{Parameswaran2012}%
  \BibitemOpen
  \bibfield  {author} {\bibinfo {author} {\bibfnamefont {S.~A.}\ \bibnamefont
  {Parameswaran}}, \bibinfo {author} {\bibfnamefont {S.~A.}\ \bibnamefont
  {Kivelson}}, \bibinfo {author} {\bibfnamefont {E.~H.}\ \bibnamefont
  {Rezayi}}, \bibinfo {author} {\bibfnamefont {S.~H.}\ \bibnamefont {Simon}},
  \bibinfo {author} {\bibfnamefont {S.~L.}\ \bibnamefont {Sondhi}}, \ and\
  \bibinfo {author} {\bibfnamefont {B.~Z.}\ \bibnamefont {Spivak}},\ }\href
  {\doibase 10.1103/PhysRevB.85.241307} {\bibfield  {journal} {\bibinfo
  {journal} {Phys. Rev. B}\ }\textbf {\bibinfo {volume} {85}},\ \bibinfo
  {pages} {241307} (\bibinfo {year} {2012})}\BibitemShut {NoStop}%
\bibitem [{\citenamefont {Nielsen}\ \emph {et~al.}(2008)\citenamefont
  {Nielsen}, \citenamefont {Bhatt},\ and\ \citenamefont {Huse}}]{Nielsen2008}%
  \BibitemOpen
  \bibfield  {author} {\bibinfo {author} {\bibfnamefont {E.}~\bibnamefont
  {Nielsen}}, \bibinfo {author} {\bibfnamefont {R.~N.}\ \bibnamefont {Bhatt}},
  \ and\ \bibinfo {author} {\bibfnamefont {D.~A.}\ \bibnamefont {Huse}},\
  }\href {\doibase 10.1103/PhysRevB.77.054432} {\bibfield  {journal} {\bibinfo
  {journal} {Phys. Rev. B}\ }\textbf {\bibinfo {volume} {77}},\ \bibinfo
  {pages} {054432} (\bibinfo {year} {2008})}\BibitemShut {NoStop}%
\bibitem [{\citenamefont {Babaev}\ and\ \citenamefont
  {Speight}(2005)}]{Babaev2005}%
  \BibitemOpen
  \bibfield  {author} {\bibinfo {author} {\bibfnamefont {E.}~\bibnamefont
  {Babaev}}\ and\ \bibinfo {author} {\bibfnamefont {M.}~\bibnamefont
  {Speight}},\ }\href {\doibase 10.1103/PhysRevB.72.180502} {\bibfield
  {journal} {\bibinfo  {journal} {Phys. Rev. B}\ }\textbf {\bibinfo {volume}
  {72}},\ \bibinfo {pages} {180502} (\bibinfo {year} {2005})}\BibitemShut
  {NoStop}%
\bibitem [{\citenamefont {Babaev}\ \emph {et~al.}(2012)\citenamefont {Babaev},
  \citenamefont {Carlstr\"om}, \citenamefont {Garaud}, \citenamefont {Silaev},\
  and\ \citenamefont {Speight}}]{Babaev20122}%
  \BibitemOpen
  \bibfield  {author} {\bibinfo {author} {\bibfnamefont {E.}~\bibnamefont
  {Babaev}}, \bibinfo {author} {\bibfnamefont {J.}~\bibnamefont {Carlstr\"om}},
  \bibinfo {author} {\bibfnamefont {J.}~\bibnamefont {Garaud}}, \bibinfo
  {author} {\bibfnamefont {M.}~\bibnamefont {Silaev}}, \ and\ \bibinfo {author}
  {\bibfnamefont {J.~M.}\ \bibnamefont {Speight}},\ }\href {\doibase
  http://dx.doi.org/10.1016/j.physc.2012.01.002} {\bibfield  {journal}
  {\bibinfo  {journal} {Physica C: Superconductivity}\ }\textbf {\bibinfo
  {volume} {479}},\ \bibinfo {pages} {2 } (\bibinfo {year} {2012})}\BibitemShut
  {NoStop}%
\bibitem [{\citenamefont {Moshchalkov}\ \emph {et~al.}(2009)\citenamefont
  {Moshchalkov}, \citenamefont {Menghini}, \citenamefont {Nishio},
  \citenamefont {Chen}, \citenamefont {Silhanek}, \citenamefont {Dao},
  \citenamefont {Chibotaru}, \citenamefont {Zhigadlo},\ and\ \citenamefont
  {Karpinski}}]{Moshchalkov2009}%
  \BibitemOpen
  \bibfield  {author} {\bibinfo {author} {\bibfnamefont {V.}~\bibnamefont
  {Moshchalkov}}, \bibinfo {author} {\bibfnamefont {M.}~\bibnamefont
  {Menghini}}, \bibinfo {author} {\bibfnamefont {T.}~\bibnamefont {Nishio}},
  \bibinfo {author} {\bibfnamefont {Q.~H.}\ \bibnamefont {Chen}}, \bibinfo
  {author} {\bibfnamefont {A.~V.}\ \bibnamefont {Silhanek}}, \bibinfo {author}
  {\bibfnamefont {V.~H.}\ \bibnamefont {Dao}}, \bibinfo {author} {\bibfnamefont
  {L.~F.}\ \bibnamefont {Chibotaru}}, \bibinfo {author} {\bibfnamefont {N.~D.}\
  \bibnamefont {Zhigadlo}}, \ and\ \bibinfo {author} {\bibfnamefont
  {J.}~\bibnamefont {Karpinski}},\ }\href {\doibase
  10.1103/PhysRevLett.102.117001} {\bibfield  {journal} {\bibinfo  {journal}
  {Phys. Rev. Lett.}\ }\textbf {\bibinfo {volume} {102}},\ \bibinfo {pages}
  {117001} (\bibinfo {year} {2009})}\BibitemShut {NoStop}%
\bibitem [{\citenamefont {Nishio}\ \emph {et~al.}(2010)\citenamefont {Nishio},
  \citenamefont {Dao}, \citenamefont {Chen}, \citenamefont {Chibotaru},
  \citenamefont {Kadowaki},\ and\ \citenamefont {Moshchalkov}}]{Nishio2010}%
  \BibitemOpen
  \bibfield  {author} {\bibinfo {author} {\bibfnamefont {T.}~\bibnamefont
  {Nishio}}, \bibinfo {author} {\bibfnamefont {V.~H.}\ \bibnamefont {Dao}},
  \bibinfo {author} {\bibfnamefont {Q.}~\bibnamefont {Chen}}, \bibinfo {author}
  {\bibfnamefont {L.~F.}\ \bibnamefont {Chibotaru}}, \bibinfo {author}
  {\bibfnamefont {K.}~\bibnamefont {Kadowaki}}, \ and\ \bibinfo {author}
  {\bibfnamefont {V.~V.}\ \bibnamefont {Moshchalkov}},\ }\href {\doibase
  10.1103/PhysRevB.81.020506} {\bibfield  {journal} {\bibinfo  {journal} {Phys.
  Rev. B}\ }\textbf {\bibinfo {volume} {81}},\ \bibinfo {pages} {020506}
  (\bibinfo {year} {2010})}\BibitemShut {NoStop}%
\bibitem [{\citenamefont {Gutierrez}\ \emph {et~al.}(2012)\citenamefont
  {Gutierrez}, \citenamefont {Raes}, \citenamefont {Silhanek}, \citenamefont
  {Li}, \citenamefont {Zhigadlo}, \citenamefont {Karpinski}, \citenamefont
  {Tempere},\ and\ \citenamefont {Moshchalkov}}]{Gutierrez2012}%
  \BibitemOpen
  \bibfield  {author} {\bibinfo {author} {\bibfnamefont {J.}~\bibnamefont
  {Gutierrez}}, \bibinfo {author} {\bibfnamefont {B.}~\bibnamefont {Raes}},
  \bibinfo {author} {\bibfnamefont {A.~V.}\ \bibnamefont {Silhanek}}, \bibinfo
  {author} {\bibfnamefont {L.~J.}\ \bibnamefont {Li}}, \bibinfo {author}
  {\bibfnamefont {N.~D.}\ \bibnamefont {Zhigadlo}}, \bibinfo {author}
  {\bibfnamefont {J.}~\bibnamefont {Karpinski}}, \bibinfo {author}
  {\bibfnamefont {J.}~\bibnamefont {Tempere}}, \ and\ \bibinfo {author}
  {\bibfnamefont {V.~V.}\ \bibnamefont {Moshchalkov}},\ }\href {\doibase
  10.1103/PhysRevB.85.094511} {\bibfield  {journal} {\bibinfo  {journal} {Phys.
  Rev. B}\ }\textbf {\bibinfo {volume} {85}},\ \bibinfo {pages} {094511}
  (\bibinfo {year} {2012})}\BibitemShut {NoStop}%
\bibitem [{\citenamefont {Carlstr\"om}\ \emph {et~al.}(2011)\citenamefont
  {Carlstr\"om}, \citenamefont {Garaud},\ and\ \citenamefont
  {Babaev}}]{Carlstrom2011a}%
  \BibitemOpen
  \bibfield  {author} {\bibinfo {author} {\bibfnamefont {J.}~\bibnamefont
  {Carlstr\"om}}, \bibinfo {author} {\bibfnamefont {J.}~\bibnamefont {Garaud}},
  \ and\ \bibinfo {author} {\bibfnamefont {E.}~\bibnamefont {Babaev}},\ }\href
  {\doibase 10.1103/PhysRevB.84.134515} {\bibfield  {journal} {\bibinfo
  {journal} {Phys. Rev. B}\ }\textbf {\bibinfo {volume} {84}},\ \bibinfo
  {pages} {134515} (\bibinfo {year} {2011})}\BibitemShut {NoStop}%
\bibitem [{\citenamefont {Dao}\ \emph {et~al.}(2011)\citenamefont {Dao},
  \citenamefont {Chibotaru}, \citenamefont {Nishio},\ and\ \citenamefont
  {Moshchalkov}}]{Dao2011}%
  \BibitemOpen
  \bibfield  {author} {\bibinfo {author} {\bibfnamefont {V.~H.}\ \bibnamefont
  {Dao}}, \bibinfo {author} {\bibfnamefont {L.~F.}\ \bibnamefont {Chibotaru}},
  \bibinfo {author} {\bibfnamefont {T.}~\bibnamefont {Nishio}}, \ and\ \bibinfo
  {author} {\bibfnamefont {V.~V.}\ \bibnamefont {Moshchalkov}},\ }\href
  {\doibase 10.1103/PhysRevB.83.020503} {\bibfield  {journal} {\bibinfo
  {journal} {Phys. Rev. B}\ }\textbf {\bibinfo {volume} {83}},\ \bibinfo
  {pages} {020503} (\bibinfo {year} {2011})}\BibitemShut {NoStop}%
\bibitem [{\citenamefont {Drocco}\ \emph {et~al.}(2013)\citenamefont {Drocco},
  \citenamefont {Reichhardt}, \citenamefont {Reichhardt},\ and\ \citenamefont
  {Bishop}}]{Drocco2012}%
  \BibitemOpen
  \bibfield  {author} {\bibinfo {author} {\bibfnamefont {J.~A.}\ \bibnamefont
  {Drocco}}, \bibinfo {author} {\bibfnamefont {C.~J.~O.}\ \bibnamefont
  {Reichhardt}}, \bibinfo {author} {\bibfnamefont {C.}~\bibnamefont
  {Reichhardt}}, \ and\ \bibinfo {author} {\bibfnamefont {A.~R.}\ \bibnamefont
  {Bishop}},\ }\href {http://stacks.iop.org/0953-8984/25/i=34/a=345703}
  {\bibfield  {journal} {\bibinfo  {journal} {J. Phys.: Condens. Matter}\
  }\textbf {\bibinfo {volume} {25}},\ \bibinfo {pages} {345703} (\bibinfo
  {year} {2013})}\BibitemShut {NoStop}%
\bibitem [{\citenamefont {Geurts}\ \emph {et~al.}(2010)\citenamefont {Geurts},
  \citenamefont {Milo\ifmmode \check{s}\else
  \v{s}\fi{}evi\ifmmode~\acute{c}\else \'{c}\fi{}},\ and\ \citenamefont
  {Peeters}}]{geurts}%
  \BibitemOpen
  \bibfield  {author} {\bibinfo {author} {\bibfnamefont {R.}~\bibnamefont
  {Geurts}}, \bibinfo {author} {\bibfnamefont {M.~V.}\ \bibnamefont
  {Milo\ifmmode \check{s}\else \v{s}\fi{}evi\ifmmode~\acute{c}\else
  \'{c}\fi{}}}, \ and\ \bibinfo {author} {\bibfnamefont {F.~M.}\ \bibnamefont
  {Peeters}},\ }\href {\doibase 10.1103/PhysRevB.81.214514} {\bibfield
  {journal} {\bibinfo  {journal} {Phys. Rev. B}\ }\textbf {\bibinfo {volume}
  {81}},\ \bibinfo {pages} {214514} (\bibinfo {year} {2010})}\BibitemShut
  {NoStop}%
\bibitem [{\citenamefont {Edstr\"{o}m}(2013)}]{Edstrom2013}%
  \BibitemOpen
  \bibfield  {author} {\bibinfo {author} {\bibfnamefont {A.}~\bibnamefont
  {Edstr\"{o}m}},\ }\href {\doibase 10.1016/j.physc.2013.01.020} {\bibfield
  {journal} {\bibinfo  {journal} {Physica C: Superconductivity}\ }\textbf
  {\bibinfo {volume} {487}},\ \bibinfo {pages} {19} (\bibinfo {year}
  {2013})}\BibitemShut {NoStop}%
\bibitem [{\citenamefont {Meilhac}\ and\ \citenamefont
  {Destainville}(2011)}]{Meilhac2011}%
  \BibitemOpen
  \bibfield  {author} {\bibinfo {author} {\bibfnamefont {N.}~\bibnamefont
  {Meilhac}}\ and\ \bibinfo {author} {\bibfnamefont {N.}~\bibnamefont
  {Destainville}},\ }\href {\doibase 10.1021/jp1099865} {\bibfield  {journal}
  {\bibinfo  {journal} {J. Phys. Chem. B}\ }\textbf {\bibinfo {volume} {115}},\
  \bibinfo {pages} {7190} (\bibinfo {year} {2011})}\BibitemShut {NoStop}%
\bibitem [{\citenamefont {Kim}\ \emph {et~al.}(1999)\citenamefont {Kim},
  \citenamefont {Neu},\ and\ \citenamefont {Oster}}]{Kim1999}%
  \BibitemOpen
  \bibfield  {author} {\bibinfo {author} {\bibfnamefont {K.~S.}\ \bibnamefont
  {Kim}}, \bibinfo {author} {\bibfnamefont {J.~C.}\ \bibnamefont {Neu}}, \ and\
  \bibinfo {author} {\bibfnamefont {G.~F.}\ \bibnamefont {Oster}},\ }\href
  {http://stacks.iop.org/0295-5075/48/i=1/a=099} {\bibfield  {journal}
  {\bibinfo  {journal} {Europhys. Lett.}\ }\textbf {\bibinfo {volume} {48}},\
  \bibinfo {pages} {99} (\bibinfo {year} {1999})}\BibitemShut {NoStop}%
\bibitem [{\citenamefont {Sengupta}\ \emph {et~al.}(2007)\citenamefont
  {Sengupta}, \citenamefont {Sengupta},\ and\ \citenamefont
  {Menon}}]{Sengupta2007}%
  \BibitemOpen
  \bibfield  {author} {\bibinfo {author} {\bibfnamefont {A.}~\bibnamefont
  {Sengupta}}, \bibinfo {author} {\bibfnamefont {S.}~\bibnamefont {Sengupta}},
  \ and\ \bibinfo {author} {\bibfnamefont {G.~I.}\ \bibnamefont {Menon}},\
  }\href {\doibase 10.1103/PhysRevB.75.180201} {\bibfield  {journal} {\bibinfo
  {journal} {Phys. Rev. B}\ }\textbf {\bibinfo {volume} {75}},\ \bibinfo
  {pages} {180201} (\bibinfo {year} {2007})}\BibitemShut {NoStop}%
\bibitem [{\citenamefont {Sengupta}\ \emph {et~al.}(2010)\citenamefont
  {Sengupta}, \citenamefont {Sengupta},\ and\ \citenamefont
  {Menon}}]{Sengupta2010}%
  \BibitemOpen
  \bibfield  {author} {\bibinfo {author} {\bibfnamefont {A.}~\bibnamefont
  {Sengupta}}, \bibinfo {author} {\bibfnamefont {S.}~\bibnamefont {Sengupta}},
  \ and\ \bibinfo {author} {\bibfnamefont {G.~I.}\ \bibnamefont {Menon}},\
  }\href {\doibase 10.1103/PhysRevB.81.144521} {\bibfield  {journal} {\bibinfo
  {journal} {Phys. Rev. B}\ }\textbf {\bibinfo {volume} {81}},\ \bibinfo
  {pages} {144521} (\bibinfo {year} {2010})}\BibitemShut {NoStop}%
\bibitem [{\citenamefont {Metropolis}\ and\ \citenamefont
  {Ulam}(1949)}]{Metropolis1949}%
  \BibitemOpen
  \bibfield  {author} {\bibinfo {author} {\bibfnamefont {N.}~\bibnamefont
  {Metropolis}}\ and\ \bibinfo {author} {\bibfnamefont {S.}~\bibnamefont
  {Ulam}},\ }\href
  {http://amstat.tandfonline.com/doi/full/10.1080/01621459.1949.10483310}
  {\bibfield  {journal} {\bibinfo  {journal} {J. Am. Statist. Assoc.}\ }\textbf
  {\bibinfo {volume} {44}},\ \bibinfo {pages} {335} (\bibinfo {year}
  {1949})}\BibitemShut {NoStop}%
\bibitem [{\citenamefont {Swendsen}\ and\ \citenamefont
  {Wang}(1986)}]{Swendsen1986}%
  \BibitemOpen
  \bibfield  {author} {\bibinfo {author} {\bibfnamefont {R.~H.}\ \bibnamefont
  {Swendsen}}\ and\ \bibinfo {author} {\bibfnamefont {J.-S.}\ \bibnamefont
  {Wang}},\ }\href {\doibase 10.1103/PhysRevLett.57.2607} {\bibfield  {journal}
  {\bibinfo  {journal} {Phys. Rev. Lett.}\ }\textbf {\bibinfo {volume} {57}},\
  \bibinfo {pages} {2607} (\bibinfo {year} {1986})}\BibitemShut {NoStop}%
\bibitem [{\citenamefont {Earl}\ and\ \citenamefont {Deem}(2005)}]{Earl2005}%
  \BibitemOpen
  \bibfield  {author} {\bibinfo {author} {\bibfnamefont {D.~J.}\ \bibnamefont
  {Earl}}\ and\ \bibinfo {author} {\bibfnamefont {M.~W.}\ \bibnamefont
  {Deem}},\ }\href {\doibase 10.1039/B509983H} {\bibfield  {journal} {\bibinfo
  {journal} {Phys. Chem. Chem. Phys.}\ }\textbf {\bibinfo {volume} {7}},\
  \bibinfo {pages} {3910} (\bibinfo {year} {2005})}\BibitemShut {NoStop}%
\bibitem [{\citenamefont {Allen}\ and\ \citenamefont
  {Tildesley}(1991)}]{allen1991computer}%
  \BibitemOpen
  \bibfield  {author} {\bibinfo {author} {\bibfnamefont {P.}~\bibnamefont
  {Allen}}\ and\ \bibinfo {author} {\bibfnamefont {D.~J.}\ \bibnamefont
  {Tildesley}},\ }\href {http://books.google.se/books?id=p1RSnQEACAAJ} {\emph
  {\bibinfo {title} {{Computer Simulation of Liquids}}}}\ (\bibinfo
  {publisher} {Oxford University Press},\ \bibinfo {year} {1991})\BibitemShut
  {NoStop}%
\bibitem [{\citenamefont {Debenedetti}\ and\ \citenamefont
  {Stillinger}(2001)}]{Debenedetti2001}%
  \BibitemOpen
  \bibfield  {author} {\bibinfo {author} {\bibfnamefont {P.~G.}\ \bibnamefont
  {Debenedetti}}\ and\ \bibinfo {author} {\bibfnamefont {F.~H.}\ \bibnamefont
  {Stillinger}},\ }\href {\doibase 10.1038/35065704} {\bibfield  {journal}
  {\bibinfo  {journal} {Nature}\ }\textbf {\bibinfo {volume} {410}},\ \bibinfo
  {pages} {259} (\bibinfo {year} {2001})}\BibitemShut {NoStop}%
\end{thebibliography}%
\end{document}